\documentclass[twocolumn,superscriptaddress,aps,prl]{revtex4-1}
\usepackage{color}
\usepackage{amsmath}
\usepackage{amssymb}
\usepackage{graphicx}
\usepackage{subfigure}

\begin{document}

\title{Self-Organized Bistability Associated with First-Order Phase Transitions}
 
\author{Serena di Santo}
\affiliation{Departamento de Electromagnetismo y F{\'\i}sica de la
  Materia e Instituto Carlos I de F{\'\i}sica Te\'orica y
  Computacional. Universidad de Granada.  E-18071, Granada, Spain}
\affiliation{Dipartimento di Fisica e Scienza della Terra,
  Universit\`a di Parma, via G.P. Usberti, 7/A - 43124, Parma, Italy}
\affiliation{INFN, Gruppo Collegato di Parma, via G.P. Usberti, 7/A -
  43124, Parma, Italy}\author{Raffaella Burioni}
\affiliation{Dipartimento di Fisica e Scienza della Terra,
  Universit\`a di Parma, via G.P. Usberti, 7/A - 43124, Parma, Italy}
\affiliation{INFN, Gruppo Collegato di Parma, via G.P. Usberti, 7/A -
  43124, Parma, Italy} \author{Alessandro Vezzani} \affiliation{
  IMEM-CNR, Parco Area delle Scienze 37/A - 43124 Parma, Italy}
\affiliation{Dipartimento di Fisica e Scienza della Terra,
  Universit\`a di Parma, via G.P. Usberti, 7/A - 43124, Parma, Italy}
\author{Miguel A. Mu\~noz} \affiliation{Departamento de
  Electromagnetismo y F{\'\i}sica de la Materia e Instituto Carlos I
  de F{\'\i}sica Te\'orica y Computacional. Universidad de Granada.
  E-18071, Granada, Spain}

\begin{abstract}
  Self-organized criticality elucidates the conditions under which
  physical and biological systems tune themselves to the edge of a
  second-order phase transition, with scale invariance.  Motivated by
  the empirical observation of bimodal distributions of activity in
  neuroscience and other fields, we propose and analyze a theory for
  the self-organization to the point of phase-coexistence in systems
  exhibiting a first-order phase transition.  It explains the
  emergence of regular avalanches with attributes of scale-invariance
  which coexist with huge anomalous ones, with realizations in many
  fields.
 \end{abstract}

\maketitle
Multistability --understood as the existence of diverse stationary
states under a fixed set of conditions-- is ubiquitous in physics and
in biology \cite{Binder,Binney,bistability-neurons}.  Bistable
switches are a common theme in the regulation of cellular processes
such as cycles, differentiation and apoptosis \cite{Cells} and, often,
genes are expressed in huge episodic bursts interspersed with periods
of quiescence \cite{eu}.  The cerebral cortex exhibits bistability
during deep sleep, with an alternation between high or low levels of
neural activity \cite{Tsodyks,Mejias,Hidalgo}.  Real neural networks, both
\emph{in vitro} and \emph{in vivo} have been reported to exhibit
power-law distributed avalanches of activity --interpreted to be a
sign of underlying criticality-- \cite{BP}; however, when inhibitory
mechanisms are repressed or under epileptic conditions \cite{Lucilla},
very large events (beyond the expectations of criticality) appear, and
size-distributions become bimodal, suggesting some kind of underlying
bistability.

Here we are interested in spatially extended noisy systems --such as
the whole cortex or gene-expression patterns across tissues-- for
which a statistical mechanics framework is most appropriate. In this
context, bistability is tantamount to the existence of a first-order
phase transition at which two phases coexist \cite{Binney}. A
cornerstone result of equilibrium thermodynamics, the \emph{Gibbs'
  phase rule}, establishes that two phases can coexist only at a
single transition point of a one-dimensional parameter space
\cite{Binney} (see however, \cite{Telo}).  Thus, if biological systems
operate in regimes of bistability, there should exist mechanisms by
which they self-tune to the edge of a first-order phase
transition. This idea resembles the rationale behind self-organized
criticality (SOC) \cite{BTW,Bak,Jensen,Pruessner,BJP}, which explains
why critical-like phenomena are ubiquitous despite the fact that
second-order phase transitions, with their associated criticality,
power-laws and scaling, occur only at singular points of phase
spaces. SOC toymodels, such as sandpiles \cite{BTW,Manna,Oslo}),
illustrate how self-tuning to criticality may occur (see below).
Theoretical progress \cite{FES-PRE,FES-PRL,BJP,JABO1} allowed for a
rationalization of how SOC works, by relating it to a standard
second-order phase transition \cite{Binney,Haye}.

The purpose of the present Letter is to formulate a general theory of
\emph{self-organized bistability} (SOB) or self-organized phase
coexistence by extending ideas of self-organization to bistable
systems. To this end, we recapitulate existing models and theory of
SOC and modify them to describe systems exhibiting a first-order phase
transition.

\emph{Standard vs. ``facilitated'' sandpiles''.}  We start focusing on
an archetypical SOC model: the stochastic Manna sandpile \cite{Manna}.
We analyze both, its standard version and a modified one.  Sandgrains
(i.e. discrete tokens of stress or ``energy'') are progressively
injected at random sites of a spatially extended system one at each
time step (slow timescale). Whenever a certain local threshold
(e.g. $z=3$) is exceeded, the corresponding site becomes unstable and
all its sandgrains are redistributed randomly (as opposed to
deterministically in the original sandpile \footnote{The deterministic
  sandpile model \cite{BTW} has additional ``conservation laws'' being
  thus much harder to analyze.})  among its nearest neighbors,
possibly generating a cascade of activity or ``avalanche''. The
dynamics is conserving, except at the boundaries where sandgrains are
``dissipated'' \footnote{For SOC mechanisms in non-conserving systems
  see \cite{JABO1} and refs. therein. }.  When avalanches stop the
addition of grains is resumed, implementing a perfect separation of
timescales. Iteration of this slow-driving/dissipation mechanism leads
to a steady state in which avalanche sizes and durations are
distributed as $P(s)\sim s^{-\tau}$ and $P(t) \sim t^{-\tau_t}$ up to
a system-size dependent cutoff
\cite{Bak,Jensen,Moloney,Dhar06,GG,BJP}.

Early experimental attempts aimed at observing scale-invariant (SOC)
avalanches in real sandpiles did not find the expected power-law
distributions. Instead, they found anomalously large quasi-periodic
avalanches, that exceeded the expectations for large events in SOC
(see, e.g. Fig.4 in \cite{sawtooth}). The reason for this is that real
sandgrains have a tendency to keep on moving once they start doing so,
dragging other grains, and \emph{facilitating} the emergence of huge
avalanches. To mimic this effect in a highly-stylized way, we consider
the Manna sandpile and modify it with a facilitation mechanism. In
particular, we let sites that receive grains simultaneously from more
than one neighbor (e.g.  from $2$) to temporarily (one timestep)
decrease their instability threshold (e.g. to $z=1$). This type of
cooperative activation is expected to generate discontinuous
transitions \cite{Haye}.
\begin{figure}[tb]
  \centering \includegraphics[width=9cm,angle=0]{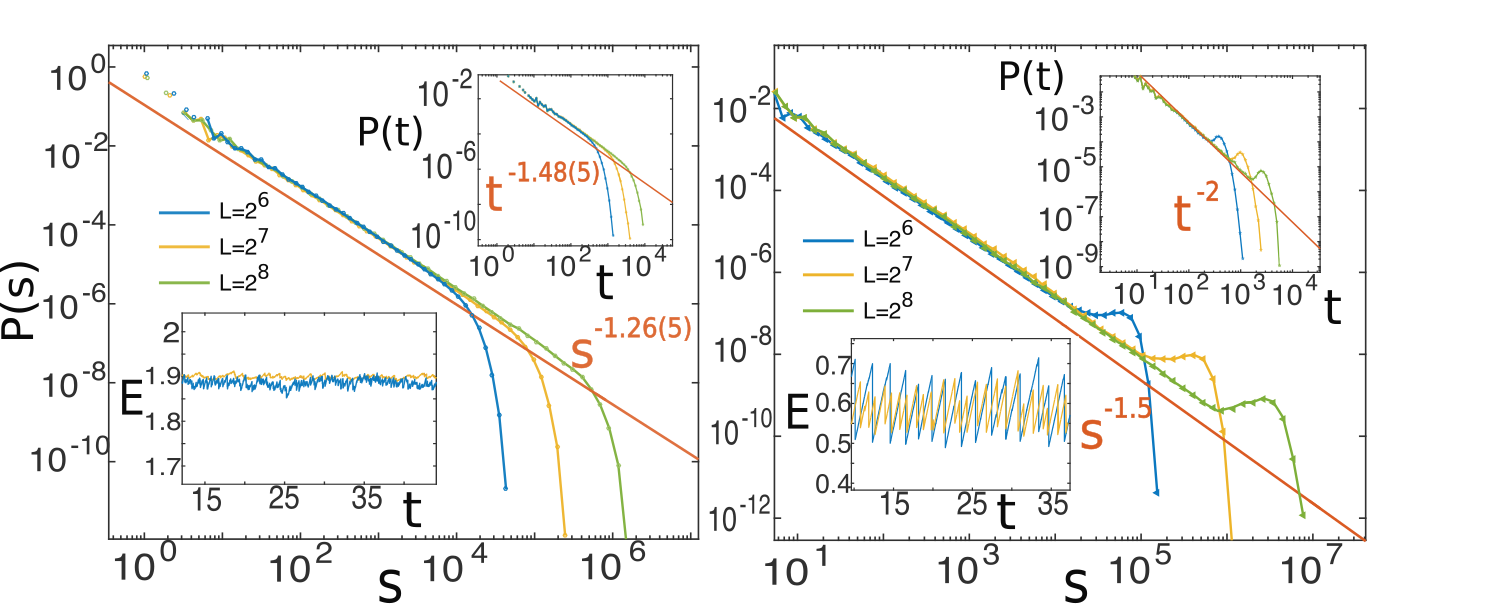} 
  \caption{Avalanche size distributions for the (LEFT) standard
    2-dimensional Manna sandpile model and the (RIGHT) facilitated
    sandpile model (time distributions for the 2 cases are shown in
    the upper insets). Observe the difference in the avalanche
    exponents, corresponding to the so-called Manna class in the
    standard (SOC) case ($\tau \approx 1.26$, $\tau_t \approx 1.48$)
    versus ($\tau \approx 3/2$, $\tau_t \approx 2$) for the
    facilitated sandpile. In the facilitated case there are bumps of
    anomalously large avalanches or ``kings'' \cite{Sornette}. The
    lower insets illustrates that``energy'' time series are much more
    sawtooth-like in the facilitated than in the SOC case owing
    to the existence of ``kings''.}
\end{figure} 
Steady-state avalanche-size distributions $P(s)$ for this facilitated
sandpile are plotted in Fig.1 for different linear system sizes, $L$.
Two facts are in blatant contrast with usual sandpile results (also
portrayed in Fig.1): {\bf(i)} the distributions are \emph{bimodal} and
consist of two different types of avalanches: ``regular ones'' and
huge avalanches or ``kings'' \cite{Sornette} --corresponding to the
bumps in the distributions-- which reverberate through the whole
system, and {\bf(ii)} regular avalanches are (nearly) power-law
distributed, but with an exponent $\tau \approx 1.5$ significantly
different from the value $\tau= 1.26(5)$ of standard sandpiles
\cite{avalanches}, \footnote{See Supplemental material at [] for a
  movie showing the alternance of both types of avalanches.}.  The
relative abundance of regular and king avalanches can be altered by
changing model details. In any case, the resulting bimodal
distributions stem from the self-organization to a state of
bistability, as will shall show by putting these findings onto a much
general framework: the theory of SOB.

\emph{SOC vs SOB: mean-field picture--} The key idea to elucidate how
SOC works consists in ``regularizing'' sandpiles by switching off slow
driving and boundary dissipation. In this way, the total amount of
sand (that we call ``energy'', $E$) becomes a conserved quantity that
can be used as a control parameter \cite{BJP,FES-PRE,FES-PRL}. In the
``{\it fixed-energy ensemble}'' the system can be either in an {\it
  active phase} (with perpetual activity) for large values of $E$, or
in an absorbing phase (where dynamics ceases) for sufficiently small
values of $E$ \cite{Haye}. Separating these two phases, there is a
critical point, $E_c$, at which a standard second-order phase
transition occurs.  In this setting, SOC is understood as a dynamical
mechanism which, by exploiting slow driving and energy dissipation at
infinitely separated timescales, self-tunes the system to $E_c$
\cite{BTW,GG,Jensen,Pruessner}). To illustrate these ideas, let us
recall how do they operate in the simplest possible mean-field
framework \cite{SOBP}.  For this, we consider, the minimal form
$\dot{\rho}(t) = a \rho - b \rho^2$ for a (mean-field) continuous
phase transition separating an absorbing phase with vanishing activity
$\rho=0$ (for $a<0$) from an active one $\rho=a/b\neq 0$ (for $a>0$);
$b>0$ is a constant (see Fig.2).  This equation is now coupled to an
additional conserved ``energy'' variable $E$ fostering the creation of
further activity, $\dot{\rho}(t) = (a + \omega E) \rho - b \rho^2$,
where $\omega>0$ is a constant. For sandpiles, $E$ represents the
total density of sandgrains while $\rho$ is the density of sites above
threshold. In the fixed-energy variant, $E$ is a conserved quantity,
and the critical point lies at $E_c=-a/\omega$. Instead, in the SOC
version $E$ is a dynamical variable, as an arbitrarily small driving
rate, $h$, and activity-dependent energy dissipation, $\epsilon$ are
switched on: $\dot{E}= h -\epsilon \rho$.  In the double limit,
$h,\epsilon \rightarrow 0$ with $h/\epsilon \rightarrow 0$ the
steady-state solution is $E=E_c$, i.e. the system self-organizes to
criticality.
\begin{figure}[tb]
  \centering \includegraphics[width=9cm,angle=0]{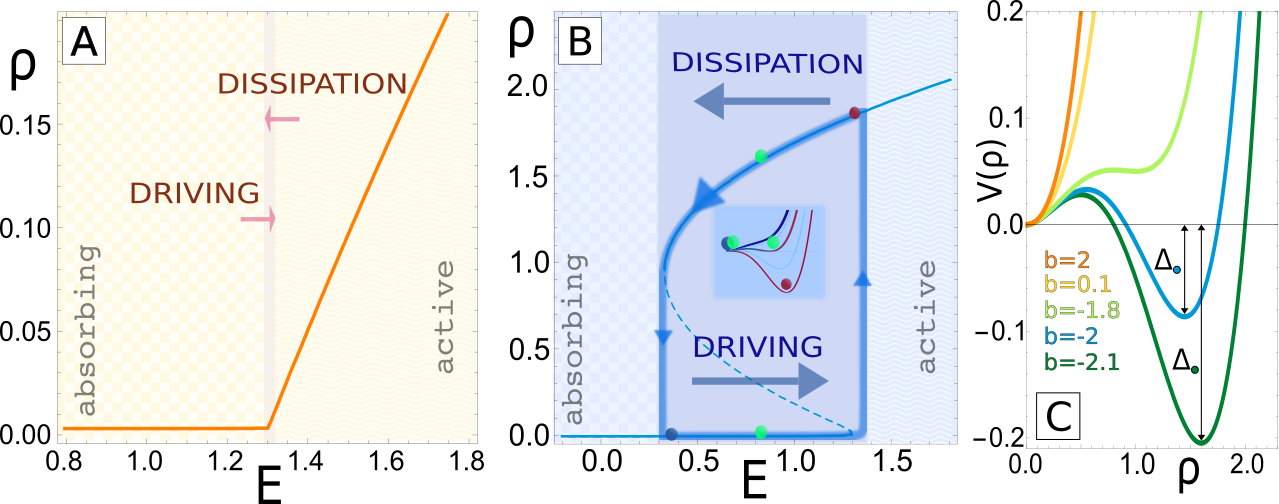} 
  \caption{Sketch of how --within mean-field theory-- the
    self-organization mechanism (alternating driving and dissipation
    at infinitely separated timescales) tunes to (A) the critical
    point of a second-order phase transition (SOC) or (B) to the
    hysteretic loop of a first-order one. In inset in (B) sketches the
    shape of the potential $V$ and the position of the minima (color
    code as in the dots of the main plot) as $E$ is changed. (C)
    Potential, $V(\rho)$ for different values of $b$, both positive
    (one minimum) and negative (two minima). For $b<0$, the potential
    depth at the active minima, $\Delta$, grows with
    $|b|$. Parameters: $a=-1.3$, $\omega=c=1$.}
\end{figure} 

To construct a mean-field theory of SOB, one needs to replace the
model showing a continuous transition, by its counterpart for a
discontinuous one: $\dot{\rho}(t) = a \rho - b \rho^2-c\rho^3$, with
$b<0$ and $c>0$ (the r.h.s. derives from the potential $V(\rho)$ shown
in Fig.2, and can be obtained from the continuous-transition case by
assuming an additional facilitation effect). Indeed, to implement a
positive feedback (facilitation) one needs to increase the $a$, in the
presence of activity, as $a \rightarrow a + \alpha \rho$, where
$\alpha$ is some constant shifting $-b$ toward larger values $b
\rightarrow -b + \alpha$.  Also, an additional cubic term is included
to avoid $\rho\rightarrow\infty$.  For the above equation, there is a
regime of bistability for the active and absorbing states, the domains
of attraction of which are separated by the \emph{spinodal} line
(dashed line in Fig.2B).  Coupling, as in SOC, this dynamics to that
of an energy field, $\dot{E}=h-\epsilon\rho$, the system follows a
limit cycle (the hysteretic loop in Fig.2): a departure from the
absorbing/active state is observed only when local stability is lost
(ending points of the spinodal line). Therefore, within the mean-field
approximation, a self-organizing mechanism identical to that of SOC
leads to cyclic bursts of activity --i.e. a sort of phase alternance
\footnote{This switching is not to be confused with stochastic
  resonance \cite{SR} which is a noise induced phenomenon.}-- rather
than to a unique point.

\emph{SOC vs SOB: beyond mean-field--} To investigate how this simple
mean-field picture changes in spatially-extended noisy systems, we
first recap the stochastic theory of SOC and, then, extend it to
first-order transitions.  The phase transition of SOC systems, in
their fixed-energy counterpart, is described, by the following set of
Langevin equations incorporating spatial coupling (diffusion) and
noise in a parsimonious way:
\begin{equation}
\begin{array}{lll}
  \partial_t \rho(\vec{x},t) = [a + \omega E({\vec{x}},t)] \rho-b\rho^{2}
  +D\nabla^2\rho+\sigma \eta({\vec{x}},t) \\ 
\partial_{t} E({\vec{x}},t) = D \nabla^2\rho({\vec{x}},t) \label{FES}
\end{array}
\end{equation}
where $\rho({\vec{x}},t)$ and $E({\vec{x}},t)$ are fields (some
dependencies on $({\vec{x}},t)$ have been omitted), $b>0$, $ D$ and
$\sigma$ are the diffusion and noise constants, respectively, and
$\eta (\vec{x},t)$ is a zero-mean multiplicative Gaussian noise with
$\langle \eta({\vec{x}},t) \eta({\vec{x'}},t) \rangle =
\rho(\vec{x},t)\delta(\vec{x}-\vec{x'}) \delta(t-t') $ imposing the
absorbing state condition. Eq.(\ref{FES}) was proposed on
phenomenological grounds \cite{FES-PRE,FES-PRL} (see also \cite{PMB})
but it can be rigorously derived from microscopic rules (using a
coherent-state path-integral representation \cite{Wiese})
\footnote{Observe that the $E(\vec{x},t)$ field is a sort of
  dynamically-generated disorder, different from the ``quenched''
  disorder appearing in other SOC-like phenomena such as Barkhaussen
  noise \cite{Sethna}.}.

The fixed-energy theory described by Eq.(\ref{FES}) exhibits a
continuous phase transition at $\bar{E}_c$ (where $\bar{E}$ is the
spatially averaged energy). More remarkably, switching on slow-driving
and boundary dissipation in Eq.(\ref{FES}) \footnote{This can be done
  in different ways; e.g. increasing both $\rho(\vec{x},t)$ and
  $E(\vec{x},t)$ at a given point by some amount ($0.1$) to create a
  new avalanche when the absorbing state has been reached and allowing
  for energy dissipation at open boundaries.} it self-organizes to
$\bar{E}^*=\bar{E}_c$. The width of the spatially-averaged energy
distribution $P(\bar{E})$ in the SOC ensemble around $\bar{E}_c$
becomes progressively smaller as system size is enlarged, ensuring
that in the thermodynamic limit the system self-organizes exactly to
its critical point \footnote{See \cite{JABO1}, and
  \cite{Fey,Fey-comment,Priezzhev} for some lingering
  controversy.}. This Langevin approach has allowed for establishing a
connection between SOC and standard non-equilibrium phase
transitions \cite{BJP,FES-PRE,FES-PRL,Pruessner}, allowing for further
computational and theoretical \cite{Dornic,cusps} understanding.

In full analogy with the mean-field case, we propose the following
equations for discontinuous transitions:
\begin{equation}
  \begin{array}{lll}
    \partial_t \rho(\vec{x},t) = [a + \omega E({\vec{x}},t)]\rho -
    b\rho^{2} - c\rho^3 
    +D\nabla^2\rho+\sigma\eta({\vec{x}},t)\\ 
\partial_{t} E({\vec{x}},t) =
  D\nabla^2\rho({\vec{x}},t),
\end{array}
\label{FES2}
\end{equation}
with $b<0$ and $c>0$.  In what follows, we vary $b$ (keeping other
parameters fixed) to explore whether diverse regimes emerge. Direct
numerical integration of Eq.(\ref{FES2}) can be performed in a very
efficient way using the split-step integration scheme of
\cite{Dornic}.  Simulations are started by either low or high
densities to enable the system to reach different homogeneous steady
states, which are separated by a spinodal line.  Results, summarized
in Fig.3, confirm that both the size of the jump and the bistability
region shrink upon reducing $|b|$ and that they shrink significantly
with respect to their mean-field values (Fig.2).  Remarkably, for
small values, e.g. $b=-0.1$, the transition becomes continuous, even
if the mean-field approximation predicts a discontinuous one. As
discussed in \cite{Eluding}, fluctuation effects typically soften the
discontinuity, shrink bistability regions, and can even alter the
order of the phase transition, leading to noise-induced criticality.
For values of $|b|$ larger than a certain (unspecified) tricritical
value $|b_T|$ the transition remains discontinuous
\cite{Tricritical}. We have also verified that there exists a point of
true phase coexistence within the bistability regime, i.e. a
\emph{Maxwell point} (defined as the value of $\bar{E}$, $E_M$, at
which a flat interface separating two halves of the system, one in
each phase, does not move on average, while, for $\bar{E}<E_M$
(resp. $\bar{E}>E_M$) the absorbing (active) phase invades the other
one; see dashed lines in Fig.3). Moreover, the observed metastability
region shrinks upon enlarging system size.
\begin{figure}[tb]
  \centering 
\includegraphics[width=9.5cm,angle=0]{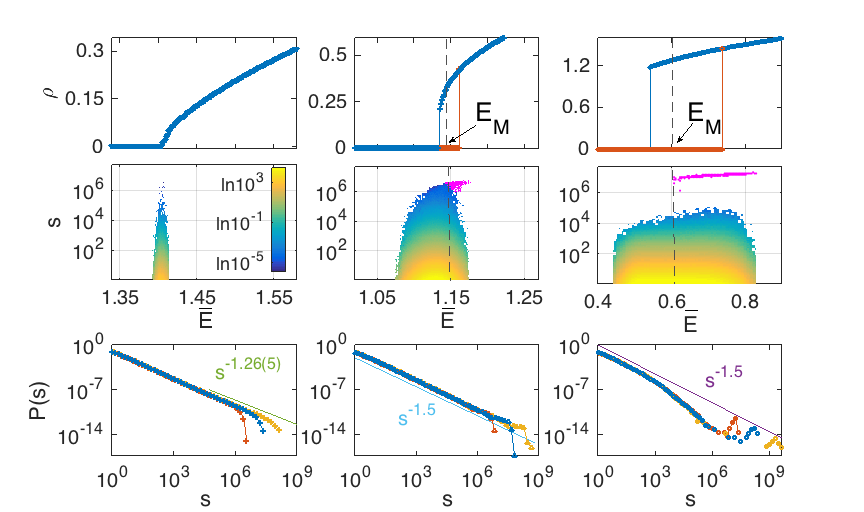} 
  \caption{The three rows show: (Upper) steady state density $\rho$ as
    a function of $E$ in the fixed-$E$ ensemble, (Central)
    color-temperature plot of the conditional size distributions
    $P(s|\bar{E})$ as a function of $E$; king avalanches plotted with a
    distinct color (magenta), and (Lower) $P(s)$ for different system
    sizes; for large $|b|$, king avalanches coexist with smaller
    ones. The three columns show three different values of $b<0$,
    ($b=-0.1$, $b=-1$ and $b=-2$, respectively) representatives of
    different regimes. System size in the first two rows is
    $L^2=2^{12}$, and $L^{2}=2^{12}, 2^{14}, 2^{16}$ in the bottom
    one. Parameter values: $a=-1.3$, $\omega=c=D=\sigma=1$.}
\end{figure}

Having characterized the fixed-energy ensemble, we now let the system
self-organize by switching on slow driving and boundary dissipation as
in SOC, and allow the system to reach its steady state.  As
illustrated in Fig.3, we observe different scenarios depending of the
value of $|b|$: \emph{(i) Noise-induced critical regime--} For
sufficiently small values of $|b|$ (such as $b=-0.1$) the transition
becomes continuous and the phenomenology is as in SOC (scale-invariant
avalanches with $\tau \approx 1.26$ and $\tau_t \approx 1.48$).
\emph{(ii) King-avalanche dominated regime--} In the opposite limit of
large values of $|b|$ (e.g. $b=-2$), we observe large peaks in $P(s)$
and $P(t)$ for large events or ``kings'', coexisting with smaller
(regular) avalanches which are exponentially truncated above a
characteristic cutoff time/size, and are responsible for large
energy-dissipation events.  \emph{(iii) Hybrid regime--} For
intermediate values of $|b|$ (e.g. $b=-1.0$), one has a situation
similar to that of the facilitated sandpile (Fig.1), in which
power-law distributed regular avalanches (with $\tau \approx 3/2$ and
$\tau_t \approx 2$) coexist with kings.  In cases \emph{(ii)} and
\emph{(iii)}, $E(t)$ exhibits characteristic sawtooh-like profiles (as
the facilitated sandpile of Fig.1) which --as revealed by the presence
of a clear peak in their power spectra (not shown)-- are
quasi-periodic, i.e. $E$ cycles between high and low values (the
larger $|b|$ the larger the excursions).  Indeed, Fig.3 (central)
shows the conditional distribution $P(s|\bar{E})$, illustrating that
avalanches can be triggered at diverse values of $\bar{E}$. However,
even if for any finite system, SOB leads to excursions all through the
bistability region, we have verified that such regions (and
excursions) shrink upon enlarging system size; thus, in the
thermodynamic limit, $\bar{E}$ self-tunes in SOB systems to a unique
point of phase coexistence --the Maxwell point-- much as in SOC
\cite{JABO1} and unlike the mean-field picture.

Let us now describe the properties of regular and king avalanches. For
regular ones, recall that right at the Maxwell point $\bar{E}=E_M$
both phases are equally stable, and thus the dynamics is as in the
so-called compact directed percolation \cite{Essam} or voter model, in
which a stable phase tries to invade an equally stable one, giving
rise to a complex dynamics at the boundaries separating both. This
type of dynamics is well-known to lead to $\tau=3/2$ and $\tau_t=2$ in
two (or larger) dimensions \cite{Essam,avalanches,Pinto}, \footnote{With the
  possibility of logarithmic corrections in two dimensions.}, so that
--as $\bar{E}$ wanders around $E_M$-- one could anticipate that
$P(s) \sim s^{-3/2}$ for regular avalanches, with some cut-off that
depends on $|b|$ (see below).

As illustrated in Fig.3, king avalanches (magenta color) can be
triggered whenever $\bar{E}$ is above the Maxwell point of the
fixed-energy diagram (Fig.3), i.e. $\bar{E} \geq E_M$ (and not only
when $\bar{E}$ reaches the limit of instability of the absorbing
state, as happens in the mean-field picture).  The reason for this
lies in the existence of a nucleation process \cite{Binder} as we
describe now. Imagine that, after driving the system, a large
fluctuation creates a large droplet of activity --of linear
size/radius $R$-- in an otherwise absorbing configuration. To
investigate the fate of such a droplet in a simple though approximate
way, we switch off noise by fixing $\sigma=0$ in Eq.(\ref{FES2}). In
this deterministic approximation, one can safely define a free energy
which has two additive contributions: one for the space integral of the
potential $V(\rho)$ (shown in Fig.2C), and a surface tension term
proportional to $D \int d\vec{x} (\nabla \rho)^2$.  When
$\bar{E}>E_M$, the potential at the active steady state ($\rho > 0$)
is negative ($\Delta<0$) and thus, deeper than that at $0$
(Fig.2C). Thus, the creation of an active droplet leads to a
competition between the gain of bulk free energy and the penalty
associated with the formation of an interface between the active and
absorbing states. Equating these two trends, one obtains a critical
radius $R_c \approx 2 D/ \Delta$ above which the bulk contribution
dominates and the droplet expands ballistically and compactly through
the whole system \cite{Binder}, giving rise to a ``king avalanche''.
This heuristic argument does not strictly apply in the presence of
(multiplicative) noise for which a free energy cannot even be
defined. However, recent analytical work has shown that the most
probable path to jump from active to inactive states in this type of
bistable noisy systems involves the creation of a critical droplet
that then expands ballistically through the system \cite{Meerson},
putting under more solid grounds our heuristic approach.
Finally, observe that the larger $|b|$ the smaller $R_c$, and the
stronger the cut-off for regular avalanches.

To visualize these effects, we have kept track of different avalanches
--both regular and kings-- and computed their averaged shape
\cite{Colaiori}; this is close to a semicircle for regular avalanches,
as correspond to random-walk like processes \cite{Colaiori}, while
kings, after a transient time, have a radically different triangular
shape (with linear growth stemming from ballistic expansion, followed
by ballistic extinction stemming from large energy
dissipation) \footnote{See Supplemental material at [...]  for an
  illustrative plot.}.

In summary, we have defined the concept of ``self-organized
bistability'' (SOB) by extending well-known ideas of self-organization
to critical points to systems exhibiting bistability and phase
coexistence and provided an explanation for the emergence of bimodal
distributions --combining aspects of scale invariance and
bistability-- as often observed in biological problems.  Our goal here
is not that of analyzing a specific example of a real system
exhibiting SOB --of which we believe there are plenty of instances--
but rather to characterize the general mechanism, much as done in
SOC. The most promising specific example to be pursued is that
provided by real neural networks (for which synaptic resources play
the role of $E$ and neural activity that of $\rho$), in which
avalanches appear to be distributed with exponents $\tau \approx 3/2$
and $\tau_t \approx 2$ \cite{BP}. These values --at odds with the
expectations of SOC in either $2$ or $3$-dimensional systems-- are
usually justified by making assumptions about the architecture of the
underlying network of connections, a hypothesis which is not always
obvious.  Furthermore, anomalously large (king) events, inconsistent
with the predictions from criticality, appear when inhibitory
mechanisms are repressed or under epileptic conditions \cite{Lucilla}
and a non-trivial temporal organization of neural avalanches
\cite{Lombardi2014} has been reported to exist.  Thus, we suggest that
it should be carefully scrutinized under which circumstances cortical
networks (which are known to have facilitation mechanisms) are not
self-organized to a critical point (SOC) --as usually considered-- but
to a region of bistability (SOB) with its concomitant mean-field like
avalanche exponents, the natural possibility of king avalanches, and a
non-trivial temporal organization.  In future work, we
shall extend our theory in a number of ways, including
self-organization in the absence of conservation laws and/or of
infinitely separated time-scales, as well as allowing for global
rather than point-like driving; these extensions will hopefully allow for
a more direct connection with biological systems.

\begin{acknowledgments}
  We are grateful to the Spanish-MINECO for financial support (under
  grant FIS2013-43201-P; FEDER funds) and to J.A. Bonachela,
  J. Hidalgo, and P. Moretti for extremely useful comments.
\end{acknowledgments}
\def\url#1{}


\end{document}